\documentclass{mem}
\usepackage{natbib}\usepackage{txfonts}\usepackage{balance}
\usepackage{graphicx}
\usepackage[a4paper]{hyperref}
\begin{document}
  
\title{
Inversion of coronal Zeeman and Hanle Observations to reconstruct
the coronal magnetic field
}

\author{ M. \,Kramar\inst{1,2}  \and B. \, Inhester\inst{1}
          }
\offprints{B. Inhester}

\institute{
  Max Planck Institute for Solar System Research
  37191 Katlenburg Lindau, Germany\\
  \email{binhest@mps.mpg.de}
\and
  The Catholic University, Physics Dept., 620 Michgan Ave,
  Washington, DC, USA
}

\authorrunning{Kramar}

\titlerunning{Inversion of coronal observations}

\abstract{
Hanle-effect observations of forbidden coronal line transitions and
recently also longitudinal Zeeman-effect measurements of coronal lines
show quantitative signatures of the weak coronal magnetic field. The
interpretation of these observations is, however, complicated by the
fact that they are the result of line-of-sight integrations through
the optically thin corona.
We study by means of simulated observations the possibility of applying
tomographic techniques in order to reconstruct the 3D magnetic field
configuration in the solar corona from these observations. The
reconstruction problem relates to a family of similar problems
termed vector tomography.
It is shown that Zeeman data and Hanle data alone obtained from vantage
points in the ecliptic plane alone are sensitive only to certain magnetic
field structures. For a full reconstruction it is necessary to combine the
longitudinal Zeeman and Hanle effect data.
}

\maketitle{}

\section{Introduction}
The coronal magnetic field is the main driving force for most plasma
processes in the inner corona. To understand the physics of the corona
a detailed knowledge of the state of the coronal field is therefore essential.
Unlike the photosphere, however, the low density and high temperature
of the coronal plasma make direct field measurements in the corona to be
rather difficult.
Conventionally, the magnetic field of the corona is therefore
estimated by means of extrapolations from its photospheric boundary
values. Until now, potential field approximations as an extrapolation
model were quite popular. The omission of all electrical currents in
the corona, however, misses an important part of coronal magnetic
field driven physics: as the potential field is the lowest energy
state of the field with respect to normal boundary conditions
\citep[e.g.][]{Sakurai:1989} it cannot account for the energy stored in the field
which is released in dynamical processes such as flares and CMEs.

With the advent of vector magnetograms from the solar surface it
became possible to employ more a sophisticated and more realistic
model assumption, the force-free condition, for the field
extrapolation. There is currently a considerable interest in
algorithms which solve this non-linear and ill-posed extrapolation
problem (see, e.g., the contribution by S. R\'egnier in this issue).
It is obvious, however, that extrapolations can only reproduce
those structures of the magnetic field which have a measure-able
imprint on the surface field. The results are deemed to become less
reliable the larger the distance from the surface.

In this paper we investigate therefore whether spectropolarimetric
coronal measurements of the magnetic field can be used for its
reconstruction. These measurements have a long tradition.
\cite{Charvin:1965} suggested more than 40 years ago to measure the
Hanle effect in the corona. One of the first to perform according
observations of the coronal green line at 530.3 nm was
\cite{Arnaud:1982}.
But full spectropolarimetric measurements of coronal lines are still
today a challenge. For a long time, the observation of the
longitudinal Zeeman effect was out of reach. Only recently
\cite{Lin:etal:2000,Lin:etal:2004} could demonstrate that full
spectropolarmetric observations in the corona can be achieved. This
success is partly due to improved instrumentation but also profits
largely from a change to infrared lines where the Zeeman split is
increased relative to the thermal line broadening. The Zeeman
observations of the Fe XIII 1075 nm line, however, still required
almost an hour of integration time.

For forbidden coronal lines with a life time much larger than the
Larmor period of the excited atom a first order interpretation of the
observed spectropolarimetric signals can be given as follows. The
Stokes V signal is proportional to the line-of-sight magnetic field
component $B_{\parallel}$ at the location where the line is emitted,
the Stokes Q and U polarisation signals indicate the orientation
$\mathbf{B}_{\perp}/B_{\perp}$ normal to the line of sight.

This picture of the line formation, however, is a strong
simplification. The physics of the emission of these forbidden lines
has been studied in great detail in recent years, e.g., by
\cite{House:1977}, \cite{Sahal-Brechot:1977} and \cite{Querfeld:1982}
and our understanding has been largely refined with respect to the
above crude interpretation.

The above crude picture, however, also a neglects the fact that the
coronal observations are line-of-sight integrals through an optically
thin medium and do not represent directly its local properties.
The goal of our study is to obtain a more sophisticated interpretation
of the observations also in this respect.
In particular, we want to investigate the whether these
observations suffice to determine a global model of the coronal
magnetic field.
Due to a lack of space we can here only briefly discuss the approach
we have adopted (chapter \ref{inversion}) and present initial results
of some of our test calculations (chapter \ref{testcalc}). Necessary
future work work is outlined in the final section (chapter \ref{outlook}).
More details can be found in \citet{Kramar:2005} and \citet{Kramar:etal:2006}.

\section{The inversion problem}
\label{inversion}

The inversion of line-of-sight observations is commonly referred to as
tomography. In many fields this is a well established technique which
solves for the distribution of an isotropic scalar quantity in a
limited region of space from line-of-sight integrals of the scalar
along a discrete number of directions. The inversion is ill-posed but
has no null space which means that all structures of the density
distribution can in principle retrieved, the quality of the
reconstruction depends on the resolution and signal to noise ratio of
the individual images taken and on the angular sampling of the image
view directions.
\begin{figure*}[t!]
\hspace*{\fill}
\resizebox{0.75\hsize}{!}{%
\includegraphics[bb=40 24 381 219,clip,width=5cm]{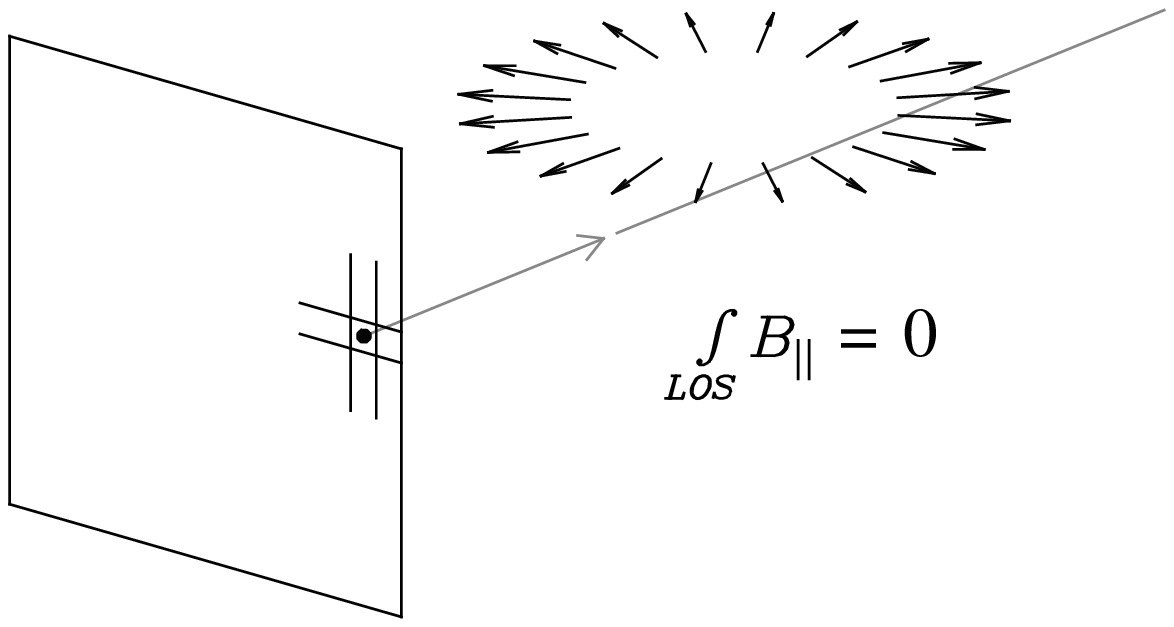}
\hspace{5mm}
\includegraphics[bb=40 24 381 219,clip,width=5cm]{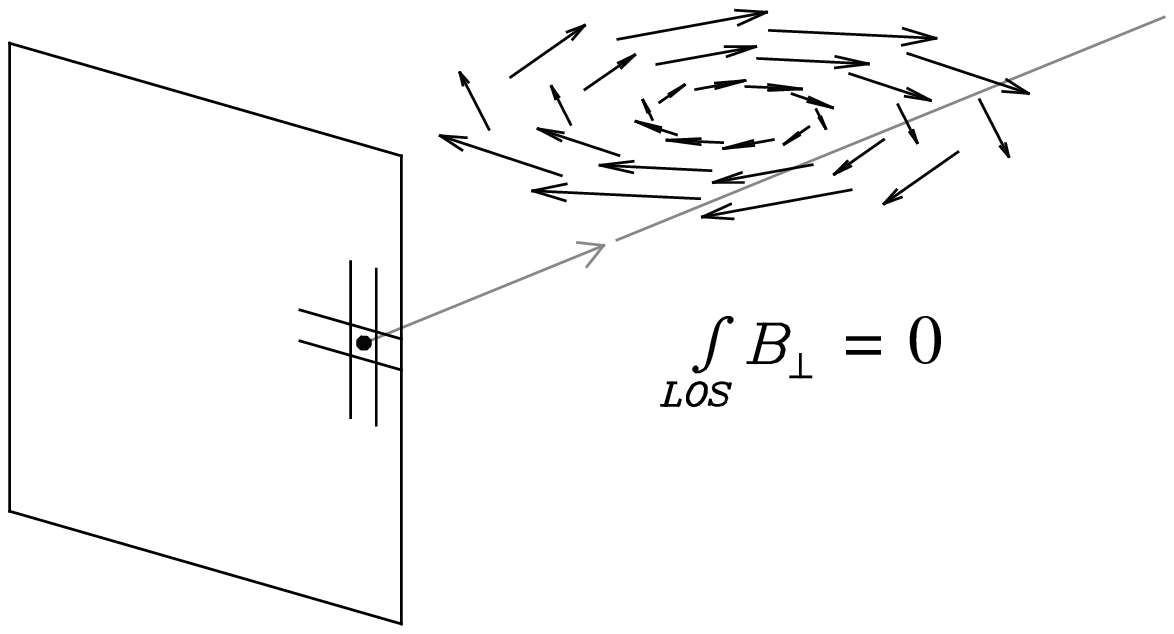}
}\hspace*{\fill}
\caption{\footnotesize Illustration of the inability to see
  div$\mathbf{B}$ and $\mathbf{e}_{LOS}\times$curl$\mathbf{B}$ from
  certain line of sight integrals}
\label{vectomo}
\end{figure*}

Tomography has also been applied to solar physics. \cite{Davila:1994}
proposed to use coronagraph observations to construct the coronal
plasma density. According calculations have been performed by
\cite{Zidowitz:1999} and \cite{Frazin:Janzen:2002}. For the
application of tomography to space observations some additional
difficulties have to be coped with which are absent in laboratory
applications. The observations are usually made with a coronagraph
with the Sun being occulted. This causes a data loss at the center of
each image which increases the condition of the inversion problem.
Observations are traditionally made from ground-based telescopes or
from space craft close to the ecliptic more or less continuously over
half a rotation period of the Sun. The tilt of the Sun's rotation axis
with respect to the ecliptic and the instationarity of coronal
structures during the time of the observation may seriously affect the
quality of the reconstruction.

The inversion problem we are faced with bears a further very fundamental
complication. We want to reconstruct a field rather than a scalar quantity
and as observations we have line of sight integrals of the field components
(or functions thereof) along and perpendicular to the view direction.
This extension to scalar tomography has been investigated in recent years
\citep[e.g.][]{Howard:1996} and has been termed vector tomography. 
If we adopt for a moment the very simplified interpretation of the
spectropolarimetric observations mentioned in the previous chapter,
for the Hanle effect we for the moment even assume that we can measure
the strength of the field in the plane of the sky,
we would observe quantities like
\begin{equation}
  \left(\begin{array}{c}
    D_V \\
    D_{(Q,U)} \\
  \end{array}\right)
       =
       \!\!\!\int\limits_\mathrm{LOS}
  \left(\begin{array}{c}
    \epsilon_V \\
    \epsilon_{(Q,U)} \\
  \end{array}\right)
       d\ell
       \sim
   \!\!\!\int\limits_\mathrm{LOS}
  \left(\begin{array}{c}
        B_{\parallel}\\
        \mathbf{B}_{\perp}
  \end{array}\right)
       d\ell
\label{LOS}\end{equation}
It is a well known result from vector tomography, that the
top integral is completely insensitive to the any divergent part
of the field $\mathrm{B}$ and the bottom integral does not respond
to the components of the curl of $\mathbf{B}$ normal to the
line of sight (see Fig.~\ref{vectomo}). Hence we have to deal with
an extended null spaces in the field we want to reconstruct from
the above measurements.

How far the problem of a finite null space also applies to real Zeeman and
Hanle effect observations is not obvious because their emissivities
are more complicated than was assumed in (\ref{LOS}). We therefore
have developed an inversion procedure to numerically test
the condition of the field inversion.
In our test computations we use the emissivity expressions
\citep{House:1977,Sahal-Brechot:1977,Querfeld:1982}
\begin{equation}
  \left(\begin{array}{c}
    \epsilon_I \\
    \epsilon_V \\
    \epsilon_Q \\
    \epsilon_U 
  \end{array}\right)
       \propto
  \left(\begin{array}{c}
    2\Sigma+ \Delta (3\cos^2\theta-1)\\
    2\Sigma\bar{g}\omega_L\cos(\theta)\\
    3\Delta\sin^2(\theta)\cos(2\alpha)\\
    3\Delta\sin^2(\theta)\sin(2\alpha)
  \end{array}\right)
\label{Stokes}\end{equation}
Here, $\Sigma$ and $\Delta$ are proportional to the ion density and
to the sum and difference, respectively, of the upper sublevel population
of the emitting electron transition, $\bar{g}$ is the effective
Land\'e factor of the transition.
The wavelength dependence of the different Stokes lines is neglected here
assuming that representative moments of the line signal were taken.
The magnetic field enters into (\ref{Stokes}) through
the field intensity in the Larmor period $\omega_L$ and
the field angles $\theta$ with respect to the line of sight
and $\alpha$, its orientation normal to the line of sight.

Our inversion procedure is based on a least square iteration between
the polarimetric observations and a forward modelling of the data
calculated from the respective line of sight integrals of (\ref{Stokes})
assuming a model field $\mathbf{B}$. The field is then successively
improved until the difference reaches the instrument's noise level.
According to the above arguments we have to expect that the forward
modelling may have null spaces and our minimisation may not yield
unambiguous results.
To stabilise our inversion, we therefore add a second term to our
expression $L(\mathbf{B})$ to be minimised
\begin{eqnarray}
  L(\mathbf{B}) =
   \sum_\mathrm{pixels}|
  \left(\begin{array}{c}
    D_V \\
    D_Q \\
    D_U \\
  \end{array}\right)^\mathrm{obs}\!\!\!
      -
  \left(\begin{array}{c}
    D_V \\
    D_Q \\
    D_U \\
  \end{array}\right)^\mathrm{sim}\!\!\!\!\!
     (\mathbf{B})\,|^2
  \nonumber \\
    +\int\limits_{\mathrm{corona}} |\mathrm{div}{\mathbf{B}}|^2 \;dv
    \hspace*{8ex}
\label{Lmin}\end{eqnarray}
This second term is an obvious constraint to the magnetic field and it
has far reaching consequences. The calculation of the divergence at the
inner coronal boundary requires the normal component of $\mathbf{B}$
on the Sun's surface as boundary condition. Hence our data set has to
be extended to include not only the coronal observations but also
the photospheric surface magnetograms.
\begin{figure*}[t!]
\hspace*{\fill}
\resizebox{0.65\hsize}{!}{%
\includegraphics[bb=24 640 334 764,clip,width=10cm]{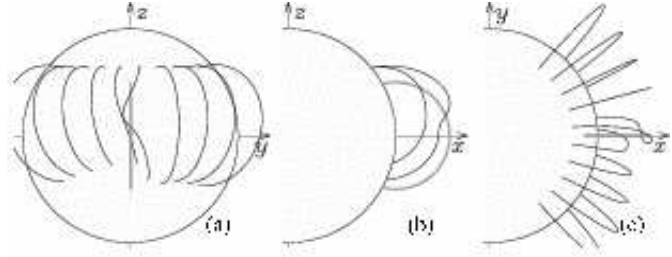}
}\hspace*{\fill}
\caption{\footnotesize
 Coronal magnetic field model 1 with a current loop in the meridional
 $x,z$ plane. Along with the current loop we also show the distorted
 field lines.}
\label{Model1}
\end{figure*}
\begin{figure*}[t!]
\resizebox{\hsize}{!}{%
\setlength{\unitlength}{1cm}
\begin{picture}(15,5)
\put(0,0){\includegraphics[bb= 23 428 151 557,clip,height=4.8cm]{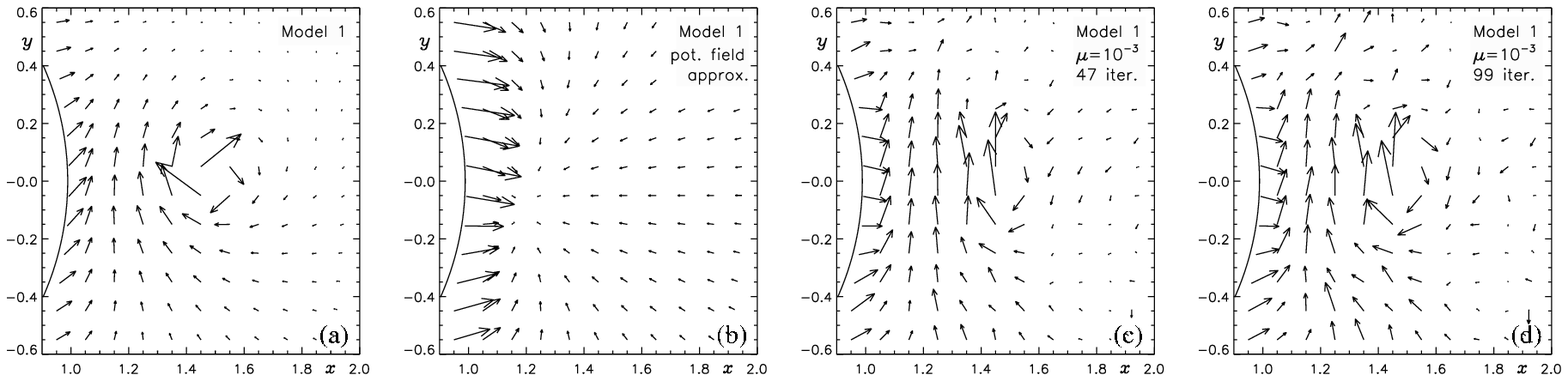}}
\put(5,0){\includegraphics[bb=425 428 553 557,clip,height=4.8cm]{Inhesterf4.eps}}
\put(10,0.1){\includegraphics[bb=80 10 581 550,clip,height=4.8cm]{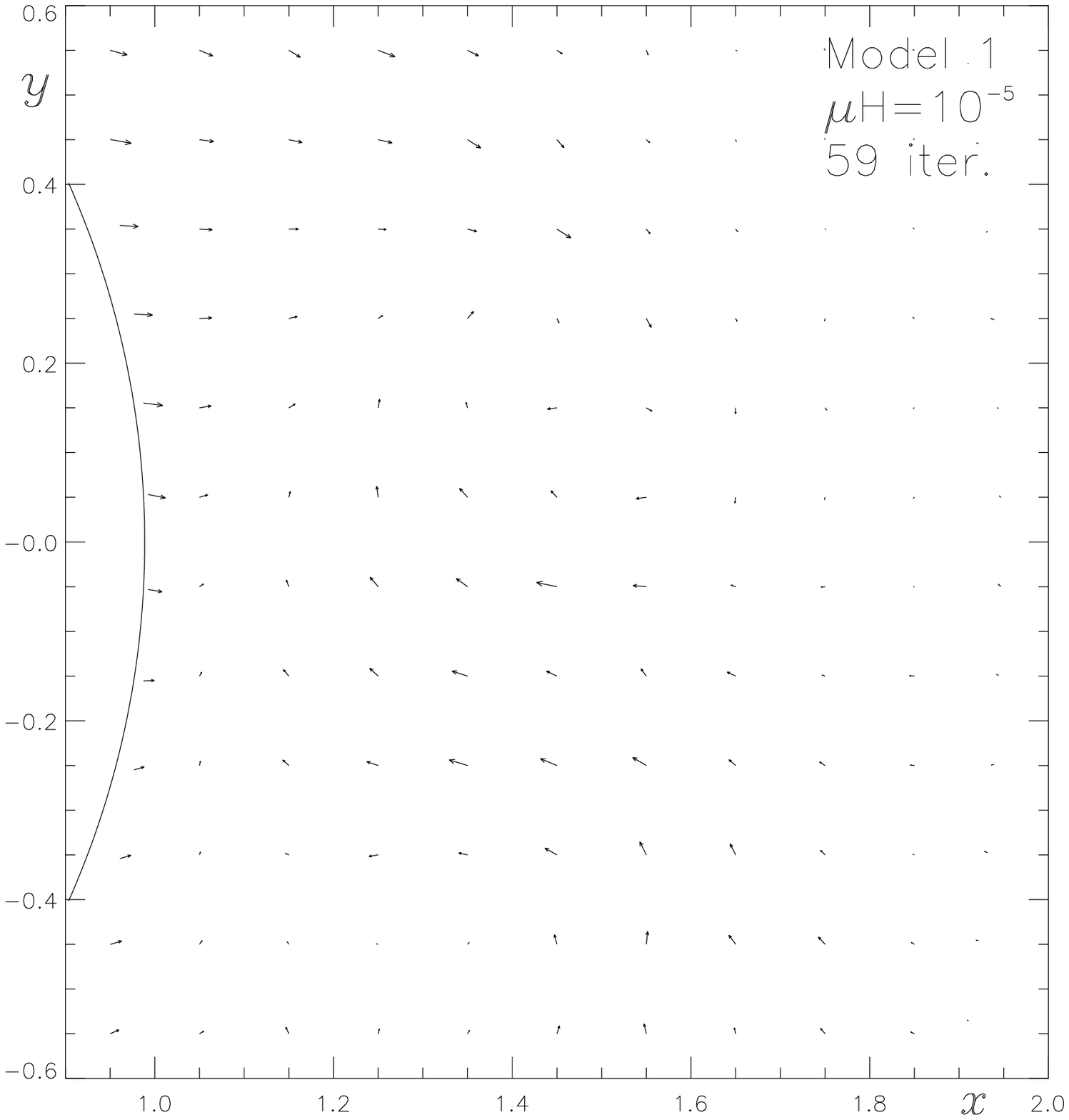}}
\end{picture}
}
\caption{\footnotesize
Field perturbations in the equatorial plane close to where the current
loop of model in Fig.~\ref{Model1} intersects the plane. From left to right
we show the original field, the reconstruction taking only account of the
Zeeman observations and of only the Hanle observations in (\ref{Lmin}).}
\label{Results1}
\end{figure*}

We consider this extension an advantage because this additional
information is merged in a natural way with our inversion procedure.
If, e.g., we had no coronal observations at all, the procedure would
only minimise the divergence term in (\ref{Lmin}). If in addition we
could insure that the total field energy $\int 
|\mathbf{B}|^2 dv$ is also minimised, the solution will be the
potential field which complies with the measured surface magnetogram.
The energy minimisation, however, is implicitly taken care of if a
proper minimisation algorithm (e.g., conjugent gradients) is chosen
along with a minimum energy initial field for the iteration.
Hence we obtain the potential field approximation of the coronal
magnetic field for free and every individual coronal observation
included in (\ref{Lmin}) drives our solution towards a more realistic,
current carrying corona.

\begin{figure*}[t!]
\hspace*{\fill}
\resizebox{0.65\hsize}{!}{%
\includegraphics[bb=33 276 551 471,clip,width=10cm]{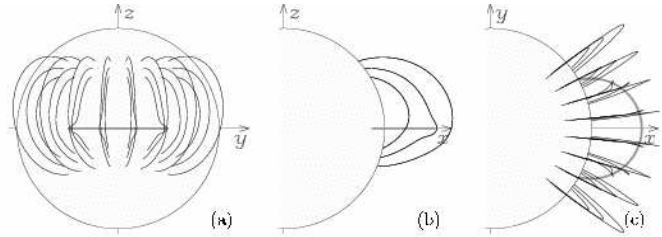}
}\hspace*{\fill}
\caption{\footnotesize
 Coronal magnetic field model 2 with a current loop in the equatorial
 $x,y$ plane. We also show the distorted field lines.}
\label{Model2}
\end{figure*}
\begin{figure*}[t!]
\resizebox{\hsize}{!}{%
\setlength{\unitlength}{1cm}
\begin{picture}(15,5)
\put(0,5){\includegraphics[bb=338  58 524 234,clip,height=4.6cm,angle=-90]%
{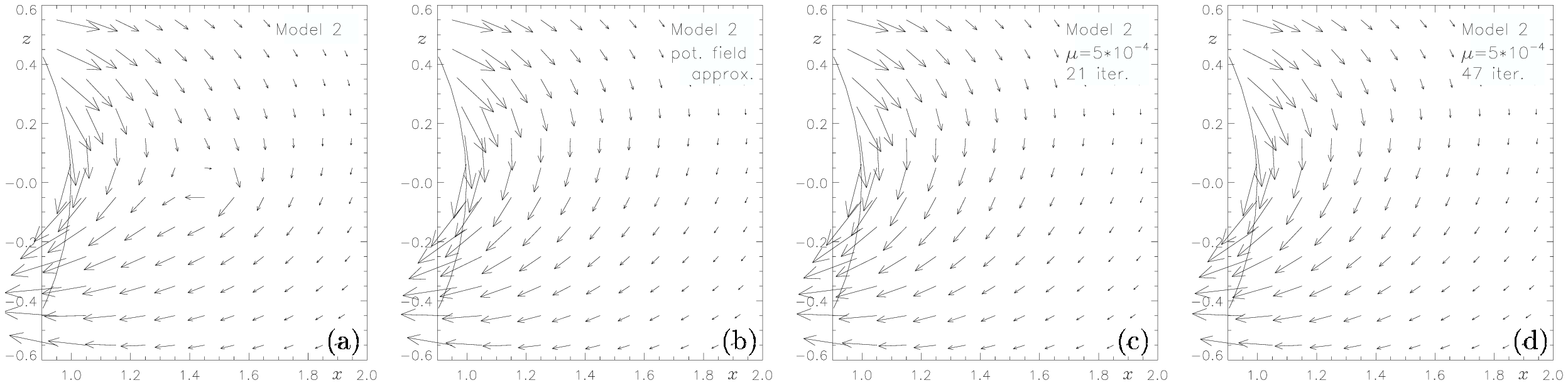}}
\put(5,5){\includegraphics[bb=338 425 524 601,clip,height=4.6cm,angle=-90]%
{Inhesterf7.eps}}
\put(10,0.1){\includegraphics[bb=300 273 544 527,clip,width=4.65cm]{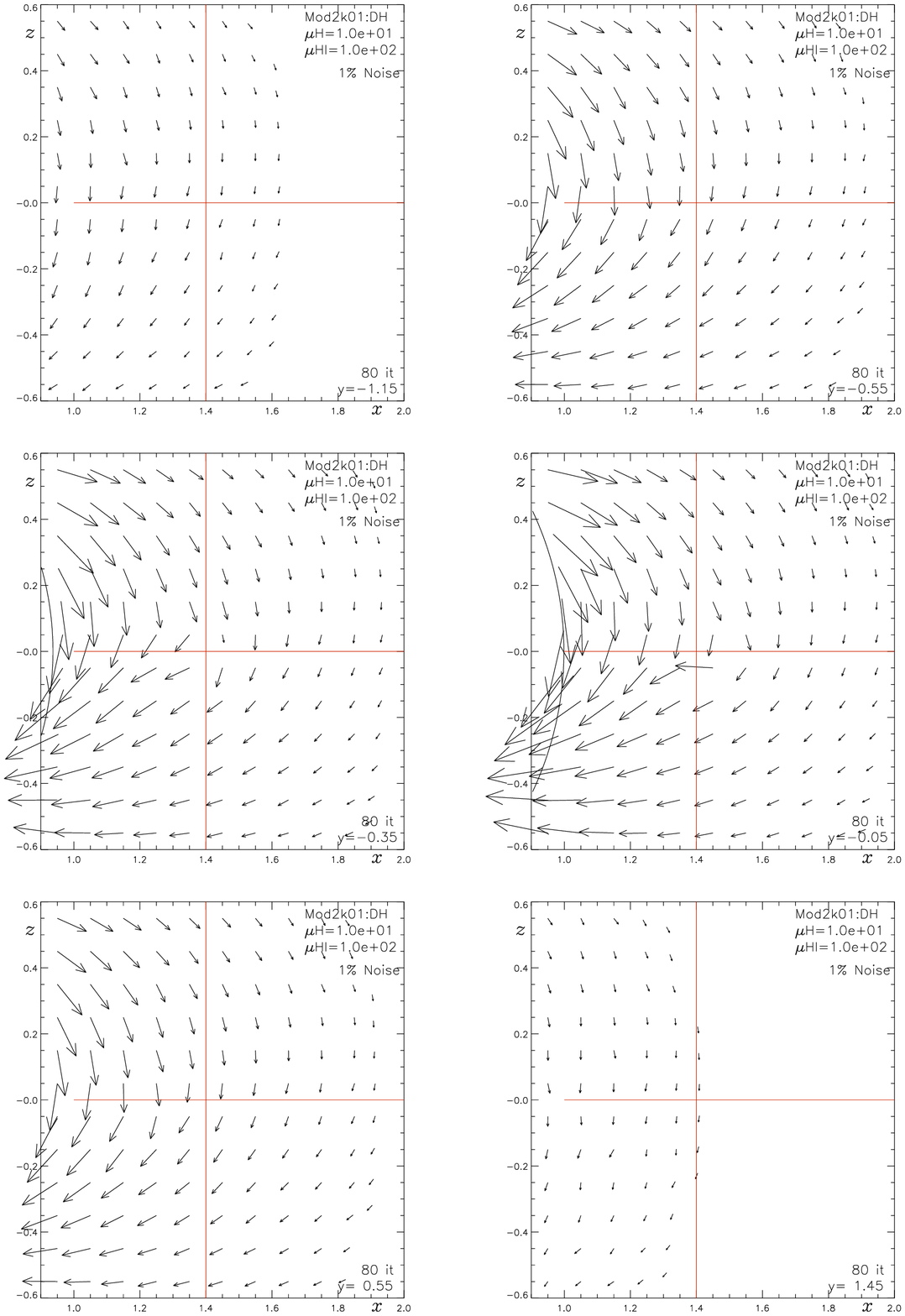}}
\end{picture}
}  
\caption{\footnotesize
Field perturbations in the meridional plane centred between the loop's foot
points and close to where the current loop of model in Fig.~\ref{Model1}
intersects the plane.
From left to right we show as in Fig.~\ref{Results1} the original field,
the reconstruction taking only account of the Zeeman observations and of
only the Hanle observations in (\ref{Lmin}).}
\label{Results2}
\end{figure*}
The problem we have to investigate can be restated much more precisely
now: is the first term in (\ref{Lmin}), i.e. the coronal
spectropolarimetric data, sufficient to resolve the magnetic field
perturbations of all current systems to be expected in the corona~?
Reversely, are there current systems which the coronal observations
do not respond to and which will consequently not appear in our
solution.
We note in passing that we could easily constrain the magnetic field
even further by adding a third term in (\ref{Lmin}) proportional to
the integral of the squared $\mathbf{j}\times\mathbf{B}$ force in the
corona. This would additionally stabilise our inversion and it is in fact
exactly the approach some popular surface field extrapolation schemes
make use of (see S. R\'egnier's contribution in this issue), except
that there the term of the coronal observations in (\ref{Lmin})
is not accounted for.

\section{Test of the inversion approach }
\label{testcalc}

In this section we report on first initial test calculations using
the approach described above.
The test consisted of the retrieval of given coronal magnetic field
configurations from simulated observations. The coronal density was
assumed to fall off radially according to a classical power law
\citep{Newkirk:1970} and assumed was known for the field inversion
because it can, at least approximately, be obtained from a scalar
tomography inversion of the line intensity.

In Fig.~\ref{Model1} we show the first model we tested. It consists of
a dipole field with an isolated current loop across the equator
confined on a meridional plane. The simulated data set was generated
form this model field as if observed by a space craft on an ecliptic
plane tilted by 10 degrees with respect to the $z$ axis.
The data set thus comprised images of the Stokes signals taken from 36
equidistantly distributed viewing directions with 5\% noise added to
the calculated signals intensities. 

The inversion cannot avoid to reconstruct the field in the whole
domain of the corona. This should not be a problem where the field
is a potential field and we therefore emphasise the perturbed field
region when presenting the inversion results.
Fig.~\ref{Results1} shows these field perturbations in the equatorial plane
in the immediate neighbourhood of the point where the current loop
intersects the plane.
Note that the background dipole field is normal to this plane so
that it does not show in Fig.~\ref{Results1}.
For test purposes we have here only made use of either the Zeeman
(centre panel in Fig.~\ref{Results1}) or the Hanle (right panel)
observations for the reconstruction.
For comparison, the left panel displays the perturbation of
the original field.
Obviously, the Zeeman observations yield a much better reconstruction
than the Hanle data. The reason seems to be that for this orientation
of the current, the Hanle data responds very little while the Zeeman
effect alone provides sufficient information to resolve this kind of
field perturbation.

As an alternative model, we investigated a dipole field with a current loop
in the equatorial plane. Again in Fig.~\ref{Results2} we show the original
and the reconstructed magnetic field in the $x,y$-plane normal to the current
loop and concentrated to the neighbourhood of the current intersection.
Note that now the background dipole field is superposed and the effect of
the current cannot so clearly be discerned.
It can be concluded though that now the Zeeman observations miss the field
perturbation entirely and only allow to reconstruct the background dipole
field. The Hanle observations on the other hand give a decent response
and its information included in the inversion reproduces to some extent
the effect of the current.

\section{Summary and outlook}
\label{outlook}

We have presented first results which aim to use coronal
spectropolarimetric observation to reconstruct the magnetic field of
the corona. From our findings we can state that Hanle or Zeeman
observation alone are not sufficient for the reconstruction. Instead the
Stokes V, Q and U components and in addition surface magnetograms are
necessarily required to be sensitive to the most obvious coronal
current systems.

We are confident that this data set is also sufficient to yield a
realistic coronal magnetic field model. This, however, has to be
verified in future experiments. If it turns out that the inversion is
too unstable to yield satisfactory field models, we still have the
option, as mentioned above, to include a force-free constraint in
our minimising function (\ref{Lmin}).

A future critical test is the dependency of the inversion output on
the noise level of the observations. This test besides its insight
into the condition of the inversion problem can also provide us with
an estimate of the maximum tolerable signal to noise ratio. This way
we may also optimise the tradeoff between image noise and angular
resolution since the Zeeman observations require long integration
times. We are confident that this information could be quite
helpful to improve real observations.

We are just at the beginning with our investigations of this new method
and there is quite some work ahead. In future we intent to also include the
Stokes I component in our data set and treat the particle density as
additional variable so that an a-priori density model is no more required.
In the more distant future we may also consider to take account of
the full spectral profile of the Stokes measurements and in return
relax a-priori assumptions about the coronal temperature distrubution.

\begin{acknowledgements}
The work by M.K. was supported by the International Max Planck Research School
on Physical Processes in the Solar System and Beyond. This research school is
located at the Max-Planck Institute for Solar System Research and at the
Universities G\"ottingen and Braunschweig.
\end{acknowledgements}

\bibliographystyle{aa}

\end{document}